\documentclass[screen]{acmart}

\AtBeginDocument{%
  \providecommand\BibTeX{{%
    \normalfont B\kern-0.5em{\scshape i\kern-0.25em b}\kern-0.8em\TeX}}}

\setcopyright{acmcopyright}
\copyrightyear{2023}
\acmYear{2023}
\acmDOI{XXXXXXX.XXXXXXX}

\usepackage{geometry}
\usepackage{makecell}
\usepackage{threeparttable}
\usepackage{multirow}
\begin{document}

%%
%% The "title" command has an optional parameter,
%% allowing the author to define a "short title" to be used in page headers.
\title{VGA: Vision and Graph Fused Attention Network for Rumor Detection}

%%
%% The "author" command and its associated commands are used to define
%% the authors and their affiliations.
%% Of note is the shared affiliation of the first two authors, and the
%% "authornote" and "authornotemark" commands
%% used to denote shared contribution to the research.
\author{Lin Bai}
\email{22120349@bjtu.edu.cn}
\affiliation{%
  \institution{Beijing Key Lab of Traffic Data Analysis and Mining}
  \institution{Beijing Jiaotong University}
  \city{Haidian}
  \state{Beijing}
  \country{China}
}

\author{Caiyan Jia}
\authornote{Corresponding Author.}
\email{cyjia@bjtu.edu.cn}
\affiliation{%
  \institution{Beijing Key Lab of Traffic Data Analysis and Mining}
  \institution{Beijing Jiaotong University}
  \city{Haidian}
  \state{Beijing}
  \country{China}
}

\author{Ziying Song}
\email{22110110@bjtu.edu.cn}
\affiliation{%
  \institution{Beijing Key Lab of Traffic Data Analysis and Mining}
  \institution{Beijing Jiaotong University}
  \city{Haidian}
  \state{Beijing}
  \country{China}
}

\author{Chaoqun Cui}
\email{21120341@bjtu.edu.cn}
\affiliation{%
  \institution{Beijing Key Lab of Traffic Data Analysis and Mining}
  \institution{Beijing Jiaotong University}
  \city{Haidian}
  \state{Beijing}
  \country{China}
}

%%
%% By default, the full list of authors will be used in the page
%% headers. Often, this list is too long, and will overlap
%% other information printed in the page headers. This command allows
%% the author to define a more concise list
%% of authors' names for this purpose.
\renewcommand{\shortauthors}{Lin Bai and Caiyan Jia, et al.}

%%
%% The abstract is a short summary of the work to be presented in the
%% article.
\begin{abstract}
With the development of social media, rumors have been spread broadly on social media platforms, causing great harm to society. Beside textual information, many rumors also use manipulated images or conceal textual information within images to deceive people and avoid being detected, making multimodal rumor detection be a critical problem. The majority of multimodal rumor detection methods mainly concentrate on extracting features of source claims and their corresponding images, while ignoring the comments of rumors and their propagation structures. These comments and structures imply the wisdom of crowds and are proved to be crucial to debunk rumors. Moreover, these methods usually only extract visual features in a basic manner, seldom consider tampering or textual information in images. Therefore, in this study, we propose a novel \underline{\textbf{V}}ision and \underline{\textbf{G}}raph Fused \underline{\textbf{A}}ttention Network (VGA) for rumor detection to utilize propagation structures among posts so as to obtain the crowd opinions and further explore visual tampering features, as well as the textual information hidden in images. We conduct extensive experiments on three datasets, demonstrating that VGA can effectively detect multimodal rumors and outperform state-of-the-art methods significantly.
\end{abstract}

%%
%% The code below is generated by the tool at http://dl.acm.org/ccs.cfm.
%% Please copy and paste the code instead of the example below.
%%
\begin{CCSXML}
<ccs2012>
   <concept>
       <concept_id>10002951.10003227.10003251</concept_id>
       <concept_desc>Information systems~Multimedia information systems</concept_desc>
       <concept_significance>500</concept_significance>
       </concept>
   <concept>
       <concept_id>10002951.10003227.10003351</concept_id>
       <concept_desc>Information systems~Data mining</concept_desc>
       <concept_significance>500</concept_significance>
       </concept>
   <concept>
       <concept_id>10010147.10010178</concept_id>
       <concept_desc>Computing methodologies~Artificial intelligence</concept_desc>
       <concept_significance>500</concept_significance>
       </concept>
 </ccs2012>
\end{CCSXML}

\ccsdesc[500]{Information systems~Multimedia information systems}
\ccsdesc[500]{Information systems~Data mining}
\ccsdesc[500]{Computing methodologies~Artificial intelligence}

%%
%% Keywords. The author(s) should pick words that accurately describe
%% the work being presented. Separate the keywords with commas.
\keywords{rumor detection, multimodal fusion, propagation structure, social media}%inter-modal similarity measure

%% A "teaser" image appears between the author and affiliation
%% information and the body of the document, and typically spans the
%% page.
%%\begin{teaserfigure}
 %% \includegraphics[width=\textwidth]{sampleteaser}
 %% \caption{Seattle Mariners at Spring Training, 2010.}
 %% \Description{Enjoying the baseball game from the third-base
 %% seats. Ichiro Suzuki preparing to bat.}
 %% \label{fig:teaser}
%%\end{teaserfigure}

%\received{20 February 2007}
%\received[revised]{12 March 2009}
%\received[accepted]{5 June 2009}

%%
%% This command processes the author and affiliation and title
%% information and builds the first part of the formatted document.
\maketitle

\section{Introduction}
With the rise of social media, billions of people use social media to browse news, share opinions, and interact with others in real time. Meanwhile, %for disseminating information has shifted from online text to multimedia. Additionally, rumors
more and more rumors appear in social media platforms with highly deceptive texts and images, %having a strong negative visual  impact. Such %graphic 
capturing the attention of readers and lowering their defenses. %As a result, some readers take those rumors as gospel and are highly influenced by them. 
Not only do these rumors cause panic and anxiety, but they also undermine society's credibility.
Therefore, it is an urgent need to accurately detect rumors by effectively capturing discriminative features from texts and images on social media platforms. % has become an urgent %issue.

\begin{figure}[t]
\centering
\includegraphics[width=0.75\textwidth]{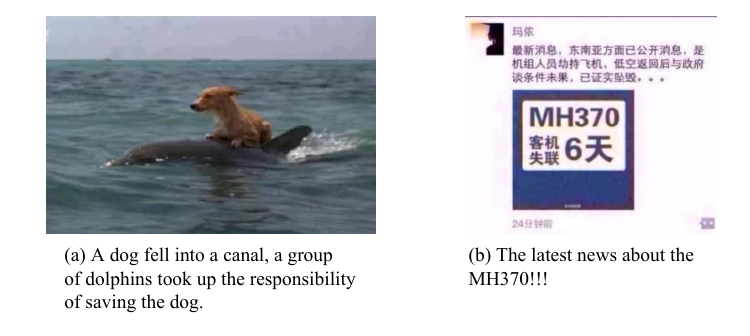}
\caption{Some indistinguishable multimodal rumors.}
\label{fig1}
\end{figure}

Traditional feature engineering-based methods and deep learning-based methods constitute the two broad categories of unimodal rumor detection methods. In traditional methods, feature extraction must be performed manually before using machine learning classifers such as Decision Tree \citep{castillo2011information} and SVM \citep{yang2012automatic}. Although traditional methods have made some progress, feature engineering is a time-consuming and laborious task and these methods have low generalization ability. On the contrary, deep learning methods can automatically learn various hidden features in a rumor dataset with low implementation costs and high model adaptability \citep{10.5555/3061053.3061153,ijcai2017p545}. Currently, Graph Neural Networks (GNNs) such as BiGCN \citep{bian2020rumor} and EBGCN \citep{wei-etal-2021-towards} are introduced to rumor detection, which learn the discriminative features of source claims and their replies modeled by propagation trees (special cases of graphs), have achieved the state-of-the-arts (SOTA) performance. This proves that learning on propagation structures of rumors should be the important part for an effective rumor detection method. 

Due to the diversification of social media, it is evident that unimodal rumor detection can no longer satisfy demand. Gradually, several multimodal rumor detection methods have been proposed. \citet{eann} proposed an Event Adversarial Neural Network (EANN) to obtain feature representations of textual and visual information, adopted adversarial training to derive event-invariant features for improving the robustness of the proposed model. \citet{safe} proposed a Similarity-Aware FakE news detection method (SAFE) to investigate the relationship between the extracted features across modalities. %t performed the similarity measure between different modalities. 
\citet{mfan} proposed a Multi-modal Feature-enhanced Attention Network (MFAN), which integrates textual, visual and social graph features in an unified framework and achieves the SOTA performance in multimodal scenario so far. All these works indicate the necessity of integrating multimodalities of rumors.

Still, the existing multimodal methods have the following limitations. (1) As depicted in Figure~\ref{fig1}(a), rumors often publish manipulated images to fabricate facts and befog the minds of individuals. The %majority of 
current methods tend to oversimplify the process of extracting image features by solely relying on VGG19 \citep{vgg} or ResNet50 \citep{resnet}, without delving deeper into the underlying visual information. 
%usually only simply use VGG19 \citep{vgg} or ResNet50 \citep{resnet} to extract image features, omitting a deeper exploration of visual information. 
Consequently, visual tampering features will be overlooked. (2) As depicted in Figure~\ref{fig1}(b), certain rumors completely conceal textual information in images, reducing the information available to a model and causing detection errors. (3) Due to the increasing camouflage ability of rumors, it is difficult to improve the performance by relying solely on source claims. %Simultaneously, 
Many methods have demonstrated the importance of propagation structures at debunking rumors, which are ignored %or underutilized 
in most of current multimodal
%rumor detection 
methods. (4) %Similarity measurement
The relationship between features of propagation structures and other modalities has not been investigated yet.

To address the aforementioned issues, we propose a novel \underline{\textbf{V}}ision and \underline{\textbf{G}}raph Fused \underline{\textbf{A}}ttention Network (VGA), % for rumor detection, 
which can better utilize propagation structures, while also exploring visual features to perform efficient rumor detection. In our model, we not only extract the images' semantic features but also obtain the images' tampering features from noise images. And the text content is extracted from images using an OCR model. Each claim's root text, reply, and reply relationship are then modeled as a rumor propagation tree to fully extract useful features including semantic features of the rumor and the crowd opinions. In addition, a similarity loss is introduced to perform the similarity measurement between features of propagation structures and other modalities. 
%two modalities. 
Furthermore, the attention mechanism is employed for an efficient multi-modal fusion in order to obtain more discriminative features for rumor detection.
%Furthermore, the attention mechanism is employed %for efficient modal fusion in order 
%to obtain effective fused features for detecting rumors.

The main contributions of the study can be summarized as follows.
\begin{itemize}
\item We propose a novel Vision and Graph Fused Attention Network (VGA) for rumor detection, which fully utilize propagation structures, meanwhile exploring visual features and the similarity relationship between graph and visual modalities.
\item We introduce SRM filter to exact noise images, which can capture tampering features of images. The attention mechanism is then used to obtain high-level representations of RGB and noise images. In addition, we adopt an OCR model to extract text content in images.
%\item By constructing each claim as an isomorphic graph and enhancing its interaction with the root node, we are able to extract propagation structure features more efficiently.
\item By constructing each claim as an isomorphic graph and enhancing its interaction with the root node through Mutually Enhanced Co-Attention, we can effectively perform rumor detection using only the visual and graph modalities, thereby replacing the need for the text modality.
\item We publish a new large multimodal rumor dataset DRWeiboMM. Empirical studies demonstrate that our VGA model can detect rumors with greater accuracy than state-of-the-art methods on all three real-world datasets.
\end{itemize}

The rest of the paper is organized as follows. Section 2 presents the related work. Next, Section 3 introduces the proposed method. The experimental results and analysis are presented in Section 4. Finally, we conclude the paper in Section 5.

\section{Related Work}

Regarding the definition of rumors, we selected a definition that is consistent with the major dictionaries and decades of social science research: "a rumor is a piece of circulating information whose truthfulness status has not been verified at the time of publication" \citep{10.1145/3161603}. Generally speaking, rumor detection can be viewed as a binary classification task comprised of Non-Rumor (NR) and False Rumor (FR), where the latter refers to rumors that have been verified to be false.
%consisting of rumors verified to be false. 
This task has the similar meaning with fake news detection.

\begin{table}[t]
  \centering
  \caption{Comparison of VGA with other methods, the notations in the table represent: textual features (T), visual features (V), graph features of isomorphic graphs (GI), graph features of heterogeneous graphs (GH), mutual enhanced co-attention (ME), similarity measurement (Sim), root enhancement (RE),  data augmentation (DA), noise images (Noise). }
  % \resizebox{0.47\textwidth}{!}{%
    \begin{tabular}{cccccccc}
    \toprule
    EANN  & T+V   & -     & -     & -     & -     & -     & - \\
    MVAE  & T+V   & -     & -     & -     & -     & -     & - \\
    SAFE  & T+V   & -     & +Sim  & -     & -     & -     & - \\
    \midrule
    EBGCN & GI  & -     & -     & -     & -     & -     & - \\
    BiGCN & GI  & -     & -     & +RE   & -     & -     & - \\
    \midrule
    MFAN  & T+V+GH & +ME   & -     & -     & -     & -     & - \\
    \midrule
    VGA   & GI+V & +ME   & +Sim  & +RE   & +DA   & +Noise & +OCR \\
    \bottomrule
    \end{tabular}%
    % }
  \label{tab0}%
\end{table}%

\subsection{Unimodal Rumor Detection}
In earlier traditional studies, the emphasis was placed on selecting and extracting essential features from rumor data. \citet{castillo2011information} extracted several features related to user posting, retweeting, original post topics, and post texts. \citet{yang2012automatic} additionally extracted two new features, posting locations and the clients used by users. \citet{kwon6729605} investigated the timing and structure of rumors to determine their features. In recent years, with the advent of deep learning. The neural network methods \citep{chen2018call,guo2018rumor,xia-etal-2020-state} gradually compensate for the disadvantages of time-consuming and laborious traditional methods.

Simultaneously, the focus of rumor detection has shifted from textual information to propagation structures %of source claims and their responses 
or visual information. As a pioneer, \citet{ma-etal-2018-rumor} generated a non-sequential tree-like rumor propagation structure based on the reply relationship between posts, achieving significant performance improvement. %achieving the first fusion of structural and semantic information.
\citet{khoo2020interpretable} proposed a variant of PLAN, Sta-PLAN, which exploited the publish-and-published relationship between posts. \citet{bian2020rumor} modeled the top-down and bottom-up reply structures as propagation and diffusion graphs and classified them using Graph Convolutional Network (GCN). \citet{mvnn} proposed a new Multi-domain Visual Neural Network framework for fusing visual information in the frequency and pixel domains to detect fake news using visual information of claims.

\subsection{Multimodal Rumor Detection}

All of the preceding methods for rumor detection use a single modality. With the proliferation of rumor types, %these methods will inevitably be constrained. In response to this phenomenon, 
multimodal methods for rumor detection have been developed. \citet{eann,metafend,mvae,spotfake} concentrated on how to use multimodal representations for classification. \citet{safe,mcnn} devoted to measure similarity between textual and visual modalities. \citet{attrnn,mkemn,carmn,emaf} focused on the alignment relationship between textual and visual modalities. But the above methods did not take propagation structures into account and did not explore deeper visual features. Despite the fact that MFAN \citep{mfan}, for the first time, exploited propagation structures by forming a heterogeneous graph of claims, their replies and user relationships, but it weakened the reply relationships with discussion nature in claims, causing MFAN to miss pointing clues to the veracity of claims.
 
To address the limitations of the above methods, we construct each claim as an isomorphic graph and enhance its interaction with the root node to extract propagation structure features more effectively while also exploring the visual features of RGB and noise images, and the similarity between modalities. Finally, we integrate propagation structure features with other modalities more effectively. Table~\ref{tab0} lists the advantages of VGA compared to other methods.

\section{Methodology}

VGA aims to further explore the use of graph (propagation structures) and visual features to improve the accuracy of rumor detection. To this end, we first apply a series of processing techniques, including noise image conversion, OCR text supplementation, node data augmentation, and root enhancement, followed by the extraction of distinctive features from each modality. To more effectively capture the correlations between the features of propagation structures and other modalities, we propose a similarity measure module. Furthermore, to enhance the fusion and alignment of the different modalities, we employ an attention mechanism. Ultimately, the fused features are utilized for classification.

%We first perform noisy image conversion, OCR text supplementation, root node enhancement, and extract the different features of each modality. To better exploit the relationship between features of propagation structures and other modalities, we propose the similarity measure module. We then use the attention mechanism for better fusion and alignment of the modalities. Finally, we use the fused features for classification. 

 \begin{figure*}[t]
\centering
\includegraphics[width=1.0\textwidth]{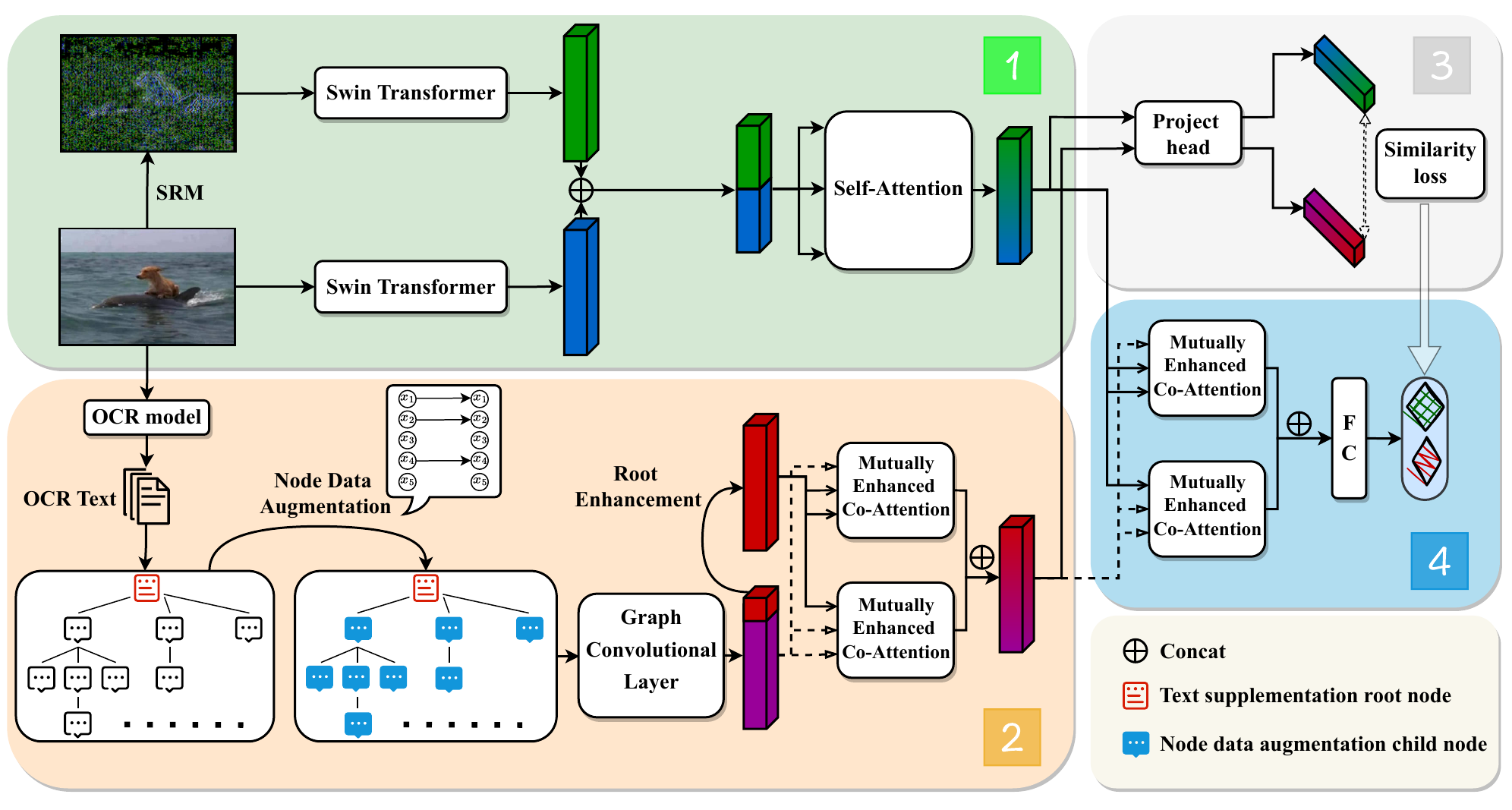}
\caption{The architecture of VGA. We first perform the conversion of noise images and image text recognition. OCR texts are used to supplement the claim root texts. Then, graph data augmentation and root enhancement are carried out. Classification of rumors is accomplished by combining image semantic features, image tampering features, and graph features in a comprehensive manner. Finally, the similarity loss is used to guide the learning of model parameters.}
\label{fig2}
\end{figure*}

Figure~\ref{fig2} illustrates its overall framework, which is comprised of four major components: visual feature extraction, graph feature extraction, inter-modal similarity measurement, multimodal fusion and classification.
%\vspace{-0.19cm}
\subsection{Problem Definition}

Let $P=\left\{p_{1}, p_{2}, \cdots, p_{m}\right\}$ represents an actual rumor dataset where $m$ is the number of claims, $p_{i}$ represents the $i$-th claim. Each claim contains two modal information, $p_{i}=\left\{G_{i}, I_{i}\right\}$ where $G_{i}$ represents the propagation structure of the $i$-th claim which characterizes "who replies or reposts to whom" relationships as a graph, and $I_{i}$ denotes the visual information of the $i$-th claim. Let graph $G_{i}=$ $<V_{i}, E_{i}>$, $\quad V_{i}=\left\{r_{i}, c_{1}^{i}, c_{2}^{i}, \cdots, c_{n_{i}}^{i}\right\}$ is the set of vertices of the graph $G_{i}$, where $r_{i}$ represents the source claim, i.e., the root text of claim $p_{i}$, $c_{i}$ represents its comments, and $n_{i}$ represents the total number of comments. $E_{i}$ is the set of edges in $G_{i}$. The adjacency matrix $A_{i} \in\{0,1\}^{\left(n_{i}+1\right) \times\left(n_{i}+1\right)}$ is used to describe the reply and repost relationships between nodes. 

Supposed that $y_{i}\in\left\{0,1\right\}$ is the label of claim $p_i$, where $y_{i}=1$ indicates False Rumor and $y_{i}=0$ indicates Non-Rumor. Our model's objective is to learn a classifier $f: P \rightarrow Y$ based on multimodal data $P$ in order to predict claim labels $Y=\{y_1,y_2,\cdots,y_m\}$.

\subsection{Visual Feature Extraction}

In this section, we separate images into RGB images and noise images, and extract their semantic and tampering features, respectively. The self-attention mechanism is then used to extract features in a higher level.

 \textbf{Image Semantic Feature Extraction.} To attract readers' attention, rumors are frequently published with visually appealing images. This results in a distinction between the semantic features of rumor RGB images and non-rumor RGB images. Based on this observation, we use Swin Transformer-Tiny \citep{swin} pre-trained in ImageNet for semantic feature extraction. We remove the Head layer of Swin Transformer-Tiny and add a fully connected layer at the end to control the dimension of the features. Thus, RGB image $I_{i}$'s semantic features $RGB_{i}$ are obtained by
 \begin{equation}
RGB_{i}=\sigma\left(W_{r g b} * \operatorname{Swin-T}\left(I_{i}\right)+b_{r g b}\right),
\label{equ1}
\end{equation}
where $RGB_{i} \in \mathbb{R}^{d / 2}$ ($d$ is the size of the dimension), $W_{rgb}$ and $b_{rgb}$ represent weights and bias in the fully connected layer respectively, $\sigma(\cdot)$ is the activation function (LeakyReLU \citep{leakyrelu} is employed here).

\textbf{Image Tampering Feature Extraction.} As Figure~\ref{fig1}(a) demonstrates, extracting image tampering information is also a critical element of rumor detection. Therefore, we introduce the image's local noise distributions to extract the tampering features by using SRM filter \citep{srm}, which enables to extract the local noise features from RGB images.

\begin{figure}[t]
\centering
\includegraphics[width=0.65\textwidth]{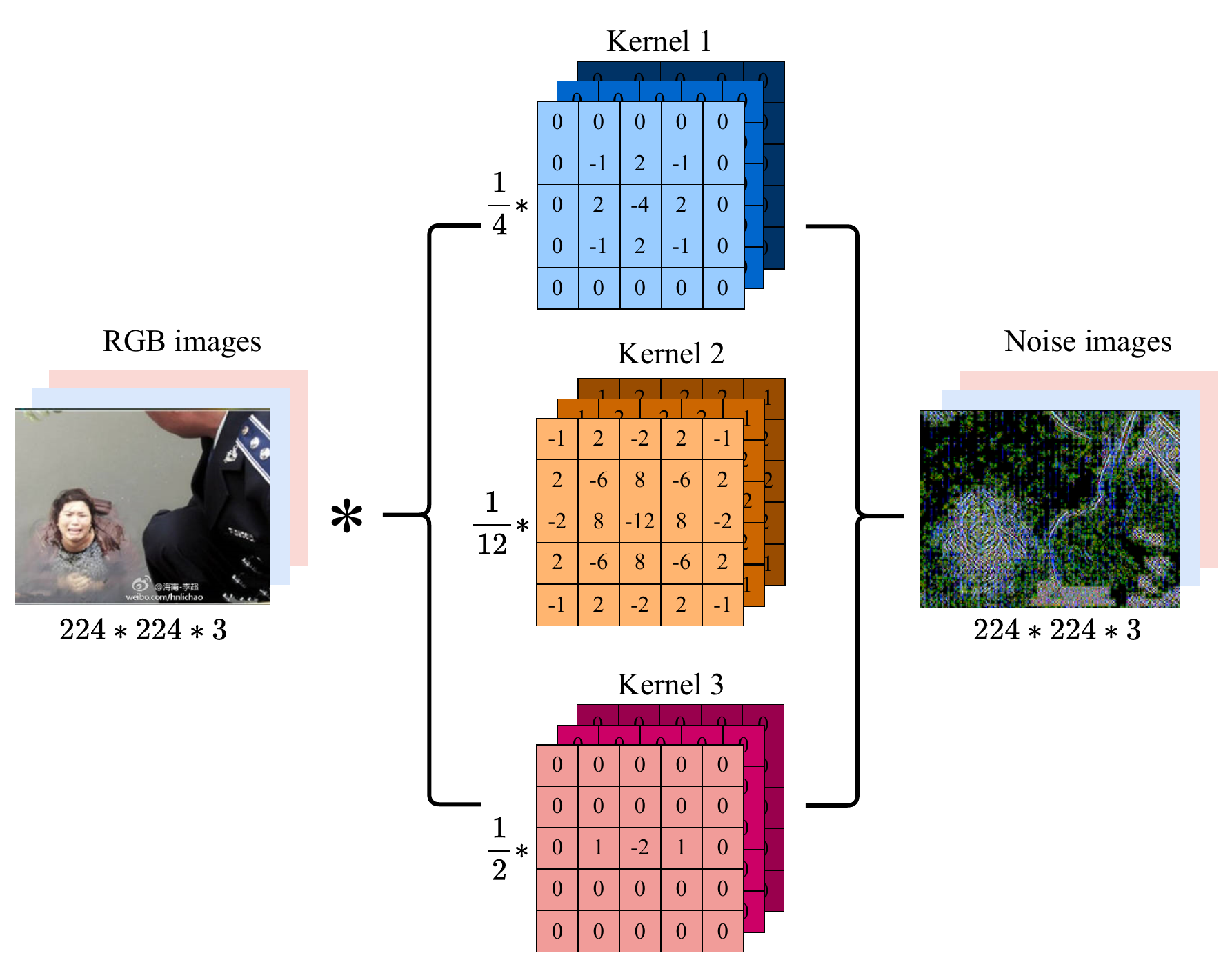}
\caption{SRM noise image conversion.}
\label{fig3}
\end{figure}

SRM is effective at differentiating tampered regions from genuine regions. It has been determined that applying all 30 SRM kernels has no appreciable effect on performance \citep{rcnn}. Therefore, we only select three kernels of size 5x5x3 for noise feature conversion, as depicted in Figure~\ref{fig3}.

Using the same network as the image semantic feature extraction module, we perform tampering feature extraction on the acquired noise image to obtain $Noise_{i}$ as follows.
\begin{equation}
Noise_{i}=\sigma\left(W_{n} * \operatorname{Swin-T}\left(I_{i}\right)+b_{n}\right),
\label{equ2}
\end{equation}
where $Noise_{i} \in \mathbb{R}^{d / 2}$, $W_{n}$ and $b_{n}$ represent weights and bias in the fully connected layer, $\sigma(\cdot)$ is also the activation function LeakyReLU. 

\textbf{Self-Attention Module.} To obtain a high-level feature representation for an image, we must enhance intra-modal feature representations using self-attention mechanism \citep{attention}. Initially, an image's semantic and tampering features are concatenated to produce $Vis_{i}$. After the following three linear transformations, $Q_{Vis}^{i}$, $K_{Vis}^{i}$, and $V_{Vis}^{i}$ are obtained, respectively.
\begin{equation}
\begin{array}{c}
Q_{V i s}^{i}=V i s_{i} W_{V i s}^{Q},\vspace{0.5ex} \\

K_{V i s}^{i}=V i s_{i} W_{V i s}^{K},\vspace{0.5ex} \\

V_{V i s}^{i}=V i s_{i} W_{V i s}^{V}.
\end{array}
\label{equ3}
\end{equation}
Where $W_{V i s}^{Q} \in \mathbb{R}^{d \times d / h}$, $W_{V i s}^{K} \in \mathbb{R}^{d \times d / h}$ and $W_{V i s}^{V} \in \mathbb{R}^{d \times d / h}$ are the trainable weight matrices, and $h$ is the number of heads. Subsequently, we can calculate the attention score of one head as follows.
\begin{equation}
\begin{array}{r}
\text {Hd}=\operatorname{Attention}\left(Q_{V i s}^{i}, K_{V i s}^{i}, V_{V i s}^{i}\right)\vspace{0.5ex}\\
=\operatorname{softmax}\left(\frac{Q_{V i s}^{i} K_{V i s}^{i}{ }^{T}}{\sqrt{\frac{d}{h}}}\right) V_{V i s}^{i}.\vspace{0.5ex}
\end{array}
\label{equ4}
\end{equation}

After calculating $h$ attention scores based on Equations~\ref{equ3} and~\ref{equ4}, the high-level representation of visual information $Vis_{i}^{\prime}$ is obtained via the subsequent operation.
\begin{equation}
Vis_{i}^{\prime}=\operatorname{Concat}\left(\text {Hd}_{1}, \ldots, \text {Hd}_{h}\right) W_{V i s}^{O},
\label{equ5}
\end{equation}
where $W_{V i s}^{O} \in \mathbb{R}^{d \times d}$ represents the weight matrix of linear transformation and $\operatorname{Concat}$ is the concatenation operation.

\begin{figure}[t]
\centering
\includegraphics[width=0.65\textwidth]{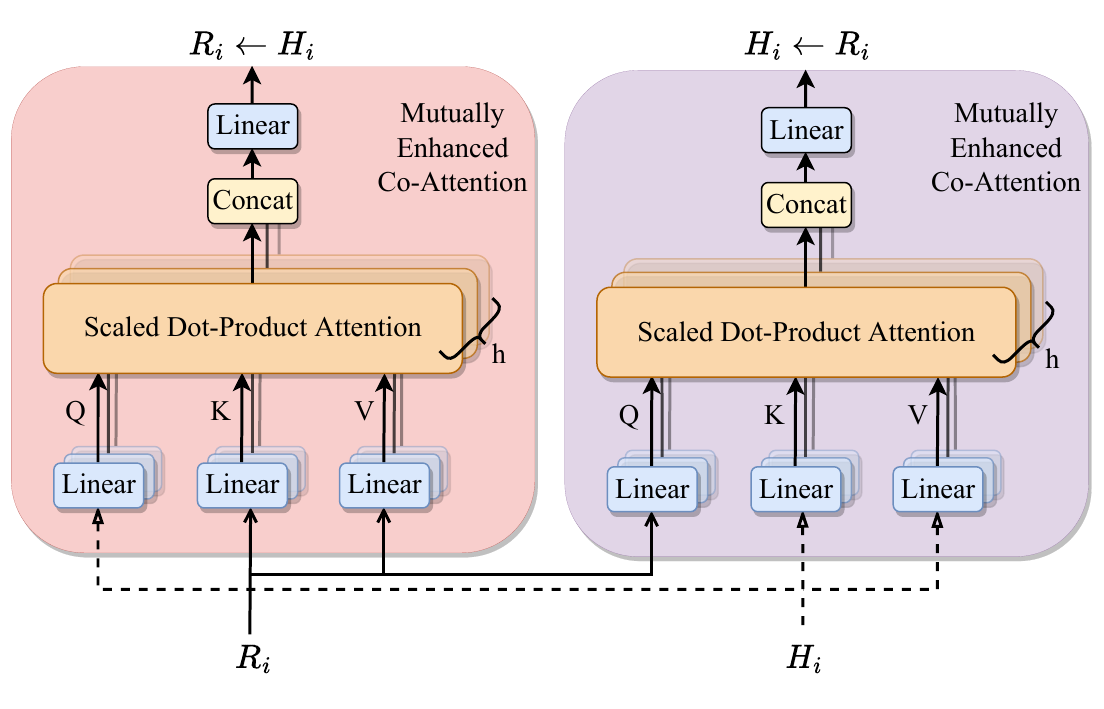}
\caption{Mutual Enhanced Co-Attention.}
\label{fig4}
\end{figure}

\subsection{Graph Feature Extraction}

\textbf{Text Supplementation and Node Data Augmentation.} Many current rumors, as depicted in Figure~\ref{fig1}(b), conceal crucial textual information within images while transforming the root node texts into a summary of the event categories or comments on the images. To extract this information, we perform text recognition on the images using OCR model\footnote{\url{https://github.com/PaddlePaddle/PaddleOCR}}. Then for each rumor $p_i$, the OCR text $T_{OCR}^{i}$ of image $I_i$ is obtained and concatenated with the root node text $r_{i}$ in order to produce the text-supplemented root node $rt_{i}$. 

%We tokenize each root node text and its comments using BERT model pre-trained on a large corpus (citation) and extract the penultimate layer representation of each token. Subsequently, all word vectors within a text are averaged to obtain the textual features of that text.

To prevent overfitting and improve the model's robustness, node data augmentation is performed on the graph data prior to feature extraction. Denoising Autoencoder \citep{da} has inspired us to add noise to the input data. In other words, the values of the nodes in the input layer are set to 0 with a predetermined probability so as to obtain the model input $V_{i}^{\prime}$ with noise. This can be interpreted as applying dropout \citep{dropout} between the input and the first layer.

\textbf{Graph Convolutional Layer.} Recently, graph neural networks, such as GCN \citep{brunagcn,kipfgcn} and Graph Attention Network (GAT) \citep{gat},   have been proposed to extract important information from graphs, used in many fields and achieved great progress. Based on our definition, since we characterize the relationships between pairs of posts of a rumor as a graph-structured data, i.e., a propagation tree, we choose GCN as our tool to extract features in the rumor graphs. %as a result of the exploration of graph data. 
%After that, \citet{kipfgcn} based on ChebyNet to take its polynomial first-order approximation,  %parameterized the graph convolution kernel using Chebyshev polynomials to reduce complexity and optimize performance,

First, we obtain the adjacency matrix $A_{i}$ and the corresponding node feature matrix $V_{i}^{\prime}$ using graph information $G_{i}=<V_{i}, E_{i}>$ of a claim $p_i$. We then perform graph convolution on them as below.
\begin{equation}
H_{i}=M\left(A_{i}, V_{i}^{\prime} ; W_{g c l}\right)=\sigma\left(\hat{A}_{i} V_{i}^{\prime} W_{g c l}\right),
\label{equ6}
\end{equation}
where $M(\cdot)$ represents the message-passing function, $W_{g c l} \in \mathbb{R}^{D \times d / 2}$ is the weight matrix, D is the text feature dimension. $\hat{A}_{i}=\widetilde{D}_{i}^{-\frac{1}{2}} \tilde{A}_{i} \widetilde{D}_{i}^{-\frac{1}{2}}$, $\tilde{A}_{i}$ is the adjacency matrix $A_i$ after adding self-join, and $\tilde{D}_{i}$ is the degree matrix.

\textbf{Root Enhancement and Mutually Enhanced Co-Attention.} As we know that all of replies and reposts in a claim are generated based on the root text also called source claim $r_i$. While most of rumor detection methods consider the root texts and their reply posts to be equivalent. We propose an attention-based root enhancement operation in order to make better use of root texts. First, the feature vector of the root node is selected from the feature matrix $H_{i}$ of all nodes, and then $n_{i}+1$ copies of the root node feature vector are aligned to obtain the matrix ${R}_{i} \in \mathbb{R}^{\left(n_{i}+1\right) \times d / 2}$, whose dimensions are the same as $H_{i}$.

Then, we use the co-attention mechanism \citep{lu2019vilbert} to mutually enhance $R_{i}$ and $H_{i}$ %and fuse their features 
(hereinafter referred to as "Mutually Enhanced Co-Attention"). This is illustrated in Figure~\ref{fig4}. However, we discover that the layer normalization and FFN operations of co-attention have a negligible effect on the performance, so we %optimize it and 
only retain multi-head attention part. To achieve interaction between the two inputs, the query matrices resulting from their linear transformation are swapped, and then a mechanism similar to self-attention is implemented below. 
\begin{equation}
\begin{array}{r}
\text {Hd}=\operatorname{Attention}\left(Q_{H}^{i}, K_{R}^{i}, V_{R}^{i}\right)\vspace{0.5ex}\\
=\operatorname{softmax}\left(\frac{Q_{H}^{i} K_{R}^{i}{ }^{T}}{\sqrt{\frac{d}{2h}}}\right) V_{R}^{i}.\vspace{0.5ex}
\end{array}
\label{equ7}
\end{equation}

Similarly, after calculating $h$ attention scores based on Equation~\ref{equ7}, the following operation is performed to obtain the root node feature matrix $R_{i} \leftarrow H_{i}$ after being enhanced by the feature matrix $H_i$ of all nodes.
\begin{equation}
R_{i} \leftarrow H_{i}=\text { Concat }\left(\text {Hd}_{1}, \ldots, \text {Hd}_{h}\right) W_{R H}^{O},
\label{equ8}
\end{equation}
where $W_{RH}^{O} \in \mathbb{R}^{d/2 \times d/2}$ represents the weight matrix of linear transformation and $\operatorname{Concat}$ is also concatenation operation.

Based on the same co-attention mechanism, we obtain the enhanced feature matrix $H_{i} \leftarrow R_{i}$ similarly. %after enhanced by the root node feature, 
%as shown in Equations~\ref{equ7} and~\ref{equ8}. %as well as 
Then, we execute the concatenation operation to obtain all node representations of the graph $G_i$.
\begin{equation}
\operatorname{Graph}_{i}=\operatorname{Concat}\left(\operatorname{R}_{i} \leftarrow H_{i}, H_{i} \leftarrow \operatorname{R}_{i}\right).
\label{equ9}
\end{equation}

The final graph representation is obtained by aggregating the information of node representations using mean pooling.
\begin{equation}
\operatorname{Graph}_{i}^{\prime}=\operatorname{Mean}\left(\operatorname{Graph}_{i}\right).
\label{equ10}
\end{equation}

\subsection{Similarity Measurement}

We find that a portion of rumors post images or memes that are completely unrelated to the root texts, which may be published for occupying more Web space and increasing readers' interests. This %phenomenon can 
introduces unnecessary noise to rumor detection methods, resulting in visual features and graph features pointing to different rumor classes, thereby degrading performance. Consequently, we propose an inter-modal similarity measure to obtain a similarity loss, %by measuring the similarity between two modalities 
and use this loss to guide the model's parameter learning.

To measure the similarity, we must ensure that the feature representations of both modalities lie in the same latent space. We first introduce Project Head \citep{simclr,graphcl} to map the feature representations $Graph_{i}^{\prime}$ and $Vis_{i}^{\prime}$ into the same space, obtain $Z_{g}^{i}$ and $Z_{v}^{i}$, respectively. Then, we compute the cosine similarity of the two vectors.
\begin{equation}
s_i=\operatorname{Sim}\left(Z_{g}^{i}, Z_{v}^{i}\right)=\frac{Z_{g}^{i T} Z_{v}^{i}}{\left\|Z_{g}^{i}\right\|\left\|Z_{v}^{i}\right\|}.
\label{equ11}
\end{equation}

In order to correlate the similarity $s_{i}$ with the claim label $y_{i}$, in other words, the low similarity between the two modalities for False Rumor and high similarity between the two modalities for Non-rumor, we map the cosine similarity $s_i$ with a Sigmoid activation function such that its value is between 0 and 1. This allows us to calculate the similarity loss based on cross-entropy loss as follows, where $\theta$ denotes all learnable parameters. %of the proposed model.
\begin{equation}
\mathcal{L}_{\text {sim }}(\theta)=-\left(y \log \left(1-s_{i}\right)+(1-y) \log s_{i}\right).
\label{equ12}
\end{equation}

\subsection{Multimodal Fusion and Classification}

We employ the mutually enhanced co-attention again to mutually enhance $Graph_{i}^{\prime}$ and $Vis_{i}^{\prime}$ of the two modalities. % in addition to feature fusion. 
In accordance with Equations like~\ref{equ7},~\ref{equ8},~\ref{equ9}, we obtain the multimodal fused representation $VG_{i}$ for a rumor $p_i$. Then, we get class prediction probability $\hat{y}_{i}$ via a fully connected layer and a softmax layer as below.
\begin{equation}
\hat{y}_{i}=\operatorname{Softmax}\left(F C\left(V G_{i}\right)\right).
\label{equ13}
\end{equation}
The classification loss is also cross-entropy loss.
\begin{equation}
\mathcal{L}_{c l s}(\theta)=-\left(y \log \hat{y}_{i}+(1-y) \log \left(1-\hat{y}_{i}\right)\right).
\label{equ14}
\end{equation}

The final objective is the sum of the classification loss and the similarity loss, where $\alpha$ is their weight.
\begin{equation}
\mathcal{L}(\theta)=\alpha \mathcal{L}_{c l s}(\theta)+(1-\alpha) \mathcal{L}_{\text {sim }}(\theta)
\label{equ15}
\end{equation}

\section{Experiments}

\subsection{Datasets}

We compare VGA with state-of-the-art methods on three datasets: WeiboCED \citep{song2019ced}, Twitter \citep{ma-etal-2017-detect} and our newly contructed dataset DRWeiboMM. \citet{song2019ced} crawls and collates weiboCED from Weibo. \citet{mfan} crawled the corresponding images of the dataset and removed the instances of data without texts or images. \citet{ma-etal-2017-detect} crawled the popular social media platform Twitter and classified the data using accuracy tags from rumor-debunking websites.
%created a rumor dataset Twitter. % from scocial platform Twitter. %and classified the data using accuracy tags from rumor-debunking websites. 
\citet{lin-etal-2022-detect}  published textual content of Twitter dataset. However, the Twitter dataset does not contain images, so we re-crawled the data and only retain the claims with both textual and visual information. Since the above two datasets are not large enough, we constructed DRWeiboMM dataset where each rumor contains both textual and visual information, covering ten years of rumors from 2012 to 2022 on Weibo. The details of these three datasets are presented in Table~\ref{tabl}. Our datasets will be available at %\url{https://anonymous.4open.science/r/MMRumorDatasets-3C17}
Github.%\footnote{\url{https://anonymous.4open.science/r/MMRumorDatasets-3C17}}. %\href{https://github.com/MeteorQ?tab=repositories}{https://github.com/MeteorQ?tab=repositories}.
%https://github.com/MeteorQ/MMRumorDatasets

\subsection{Experimental Setup}

VGA is implemented using PyTorch with the Adam optimizer to update parameters. The claim texts are initialized with BERT \citep{Bert} in dimension 768. The learning rate is set to 0.001 for WeiboCED, Twitter and 0.0005 for DRWeiboMM. The number of heads $h$ is set to 8. We conduct a grid search within [0.1,0.3,0.5,0.7,0.9] to determine the optimal $\alpha$. We divide datasets into training and validation sets in the ratio of 8:2 and use 5-fold cross-validation to train and test the models. Early stopping is used to prevent model overfitting. Patience is set to 10 with a maximum of 100 epochs trained. We report the average results of models' accuracy and F1 score, as well as the improvement of VGA over the best-performing baselines.
%, precision, recall,

\begin{table}[t]
  \centering
  \setlength{\tabcolsep}{13pt}
  \caption{Statistic of three datasets.}
    \begin{tabular}{cccc}
    \toprule
    Statistic & WeiboCED & DRWeiboMM & Twitter \\
    \midrule
    claims & 1467  & 3202  & 554 \\
    images & 1467  & 3202  & 554 \\
    false rumors & 590   & 1331  & 237 \\
    non-rumors & 877   & 1871  & 317 \\
    total posts & 583,091 & 168,244 & 27,855 \\
    avg nodes & 397   & 53    & 50 \\
    \bottomrule
    \end{tabular}%
    \label{tabl}

\end{table}%

\subsection{Baselines}
We compare our model with both graph-based and multimodal SOTA rumor detection methods. 

\textbf{Graph-based Methods.} 
\begin{itemize}

\item\textbf{BiGCN} \citep{bian2020rumor}: a bi-directional graph model that explores both top-down and bottom-up features on propagation structures of rumors. 
\item\textbf{EBGCN} \citep{wei-etal-2021-towards}: a Bayesian approach that adaptively reconsiders the reliability of potential relationship of each edge in BiGCN. 

\end{itemize}
%In addition, EBGCN applies a new edge-consistent training framework to optimize the model by performing consistency on relations.

\textbf{Multimodal Methods.} 
\begin{itemize}
\item \textbf{EANN} \citep{eann}: it uses an event discriminator to measure the dissimilarities among different events and further learns the event invariant features which can generalize well for the newly emerged events. 
\item \textbf{MVAE} \citep{mvae}: it uses a multimodal variational autoencoder trained jointly with a fake news detector to detect if a post is fake or not. 
\item \textbf{SAFE} 
\citep{safe}: it converts images to texts as a measure of similarity between the two modalities to determine the veracity of claims. 
\item \textbf{MFAN} \citep{mfan}: it combines three modalities: texts, images, and heterogeneous graphs characterizing the relationships among users, posts and images to classify claims.

\end{itemize}
\begin{comment}
    \begin{table*}[htbp]
  \centering
  \setlength{\tabcolsep}{8pt}
  %\small
  \caption{Ablation experiment results.}
  \scalebox{1.0}{
    \begin{tabular}{c|cccc|cccc|cccc}
    \toprule
    \multicolumn{1}{c|}{}&\multicolumn{4}{c|}{WeiboCED}         & \multicolumn{4}{c|}{DRWeiboMM} & \multicolumn{4}{c}{Twitter} \\
    \midrule
    Method & Acc. & Pre. & Rec. & F1. & Acc. & Pre. & Rec. & F1. & Acc. & Pre. & Rec. & F1. \\
    \midrule
    VGA w/o Sim  & 93.48 &    93.31   &    93.45   &   92.94    & 86.24 & 86.88 & 85.08 & 85.02 & 83.78 & 84.03 & 82.47 & 82.10 \\
    VGA w/o RE  & 94.67 & 93.94      &    93.88   &  93.96     & 86.80 & 86.99 & 86.04 & 85.72 & 84.93 & 85.52 & 83.85 & 83.56 \\
    VGA w/o DA  & 92.08 &     92.04  &     91.52  &      91.30 & 87.23 & 87.59 & 86.15 & 86.04 & 83.16 & 83.88  & 81.41  & 81.21 \\
    VGA w/o Noise & 93.93 & 93.81 & 93.82 & 93.36 & 88.16 & 88.65 & 87.05 & 87.07 & 84.69 & 85.29 & 83.28 & 82.96 \\
    VGA w/o OCR & 93.92 & 93.67 & 93.77 & 93.35 & 87.68 & 88.39 & 86.59 & 86.57 & 84.63 & 85.71 & 83.47 & 83.24 \\
    VGA & \textbf{95.07} & \textbf{94.75} & \textbf{95.14} & \textbf{94.64} & \textbf{88.37} & \textbf{88.71} & \textbf{87.47} & \textbf{87.39} & \textbf{85.87} & \textbf{86.85} & \textbf{84.88} & \textbf{84.61} \\
    \bottomrule
    \end{tabular}%
    }
  \label{tab3}
\end{table*}%
\end{comment}

\begin{table*}[htbp]
  \centering
  % \setlength{\tabcolsep}{8pt}
  %\small
  \caption{Performance comparison with SOTA rumor detection methods on three real-world datasets.}
  \scalebox{1.0}{
  \begin{threeparttable}
    \begin{tabular}{cc|cc|cc|cc}
    \toprule
    \multirow{2}[2]{*}{Method} & \multirow{2}[2]{*}{Modality\tnote{*}} &\multicolumn{2}{c|}{WeiboCED}& \multicolumn{2}{c|}{DRWeiboMM} & \multicolumn{2}{c}{Twitter} \\
    %\midrule
    \cmidrule{3-8}          & & \makecell[c]{Accuracy} &  \makecell[c]{F1 Score} & \makecell[c]{Accuracy}  & \makecell[c]{F1 Score} & \makecell[c]{Accuracy} &  \makecell[c]{F1 Score} \\
    \midrule
    EANN  & T+V & 80.96   &   79.87    & 76.78  & 75.23 & 78.18 &  76.89 \\
    MVAE  & T+V &  71.67    &  70.34     & 71.70  & 64.60 & 71.81  & 69.71 \\
    SAFE  & T+V &  84.95   &      84.96 & 81.74  & 77.28 & 71.71   & 68.00 \\
    \midrule
    EBGCN & GI &  83.14  & 81.45 & 81.67  & 80.99 & 82.90  & 81.01 \\
    BiGCN & GI &  88.63  & 87.83 & 82.57  & 81.56 & \underline{82.99}  & \underline{81.92} \\
    \midrule
    MFAN  & T+V+GH &  \underline{88.95}  & \underline{88.33} & \underline{84.87}  & \underline{83.92} & 80.17  & 72.35 \\
    \midrule
    VGA & GI+V &  \textbf{95.17} (+6.22)\tnote{**}  & \textbf{94.78} (+6.45) & \textbf{88.37} (+3.50)  & \textbf{87.39} (+3.47) & \textbf{85.87} (+2.88) & \textbf{84.61} (+2.69) \\
    \bottomrule
    \end{tabular}%
    \begin{tablenotes}   
        \footnotesize               % 添加命令
        \item * T means textual features, V means visual features, GI means graph features of isomorphic graphs and GH means graph features of heterogeneous graphs.   
        \item ** The improvement of VGA over the best-performing baselines (underlined) mentioned above.     
      \end{tablenotes} 
    \end{threeparttable}
    }

  \label{tab2}
\end{table*}% /

\subsection{Results and Discussion}

The experimental outcomes are presented in Table~\ref{tab2}. On all three datasets, our model performs significantly better than the other baselines. EANN outperforms MVAE. Whereas, SAFE outperforms the other two methods EANN and MVAE on Chinese datasets, demonstrating the significance of multimodal similarity measures. Its poor performance on Twitter could be attributable to the size of the dataset.
%which is easy to cause overfitting. 
Twitter's limited data quantity causes overfitting and limited generalization of SAFE. 
In graph-based methods, both EBGCN and BiGCN outperform or are comparable to the multimodal (textual + visual) methods, indicating the strong influence of propagation structures on rumor detection. Moreover, MFAN outperforms all methods except VGA in Chinese datasets, indicating that propagation structures play a positive role in multimodal rumor detection. However, the poor performance of MFAN on Twitter demonstrates that MFAN is also highly dependent on the data quantity and highly sensitive to the number of nodes in heterogeneous graphs that MFAN construct. In our experiments, we have discovered that because MFAN's heterogeneous graphs require a large amount of memory usage, it discard nodes with short text lengths, resulting in certain information loss. %Therefore, MFAN did not perfectly utilize spread structures. 
Our VGA achieves the best performance on all three datasets, demonstrating the efficacy of VGA for deeper exploration of graph and visual information, as well as its ability to achieve the best performance with excellent generalization on small datasets.

\subsection{Ablation Analysis}

\subsubsection{Validity of Each Module}
In order to verify the validity of each module, several variants of VGA have been designed. %and compared as follows: 
They include: 
\begin{itemize}
    \item \textbf{VGA w/o Sim}: a variant of VGA with similarity measurement module removed.
    \item \textbf{VGA w/o RE}: a variant of VGA with root enhancement module removed.
    \item \textbf{VGA w/o DA}: a variant of VGA with node data augmentation module removed.
    \item \textbf{VGA w/o Noise}: a variant of VGA with noise image conversion and feature extraction module removed.
    \item \textbf{VGA w/o OCR}: a variant of VGA with the image text extraction operation removed.
\end{itemize}

\begin{comment}
    \begin{table}[htbp]
  \centering
  \setlength{\tabcolsep}{10pt}
  \caption{Ablation experiment results.}  
    \begin{tabular}{cccc}
    \toprule
    Method & WeiboCED & DRWeiboMM & Twitter \\
    \midrule
    w/o Sim & 93.48 & 86.24 & 83.78 \\
    w/o RE & 94.67 & 86.80 & 84.93 \\
    w/o DA & 92.08 & 87.23 & 83.16 \\
    w/o Noise & 93.93 & 88.16 & 84.69 \\
    w/o OCR & 93.92 & 87.68 & 84.63 \\
    VGA   & \textbf{95.07} & \textbf{88.37} & \textbf{85.87} \\
    \bottomrule
    \end{tabular}%
  \label{tab3}
\end{table}%
\end{comment}

\begin{table*}[htbp]
  \centering
  % \setlength{\tabcolsep}{13pt}
  %\small
  \caption{Ablation experiment results on three real-world datasets.}
  \scalebox{1.0}{
  
  \begin{threeparttable}
    \begin{tabular}{c|cc|cc|cc}
    \toprule
    \multirow{2}[2]{*}{Method}  &\multicolumn{2}{c|}{WeiboCED}         & \multicolumn{2}{c|}{DRWeiboMM} & \multicolumn{2}{c}{Twitter} \\
     \cmidrule{2-7}  & \makecell[c]{Accuracy} &  \makecell[c]{F1 Score} & \makecell[c]{Accuracy}  & \makecell[c]{F1 Score} & \makecell[c]{Accuracy} &  \makecell[c]{F1 Score} \\
    \midrule
    VGA w/o Sim  & 93.48 (-1.69)\tnote{*}&92.94 (-1.84)& 86.24 (-2.13)& 85.02 (-2.37) & 83.78 (-2.09)  & 82.10 (-2.51) \\
    VGA w/o RE  & 94.67 (-0.50)    &  93.96 (-0.82)     & 86.80 (-1.57)  & 85.72 (-1.67) & 84.93 (-0.94)  & 83.56 (-1.05) \\
    VGA w/o DA  & 92.08 (-3.09)   &      91.30 (-3.48) & 87.23 (-1.14)  & 86.04 (-1.35) & 83.16 (-2.71)   & 81.21 (-3.40) \\
    VGA w/o Noise & 93.93 (-1.24) & 93.36 (-1.42) & 88.16 (-0.21) & 87.07 (-0.32) & 84.69 (-1.18)  & 82.96 (-1.65) \\
    VGA w/o OCR & 93.92 (-1.25)  & 93.35 (-1.43) & 87.68 (-0.69)  & 86.57 (-0.82) & 84.63 (-1.24)  & 83.24 (-1.37) \\
    VGA & \textbf{95.17}  & \textbf{94.78} & \textbf{88.37}  & \textbf{87.39} & \textbf{85.87}  & \textbf{84.61} \\
    \bottomrule
    \end{tabular}%
        \begin{tablenotes}   
        \footnotesize               % 添加命令  
        \item * Performance gap between VGA variants and VGA.     
      \end{tablenotes} 
    \end{threeparttable}
    }
  \label{tab3}
\end{table*}%

These variants are compared with the original VGA in Table \ref{tab3}. The experimental results demonstrate that the similarity measurement module significantly improves model performance across all three datasets, highlighting the significance of correlations between the features of propagation structures and other modalities for rumor detection. The node data augmentation module used in VGA also works well to prevent overfitting and enhance the model's robustness.
%both similarity measure and graph data augmentation modules have significant effects. 
The root enhancement module has less impact than the above two modules and is less powerful in WeiboCED dataset where the average number of nodes is larger. The positive effects of noise image feature extraction and OCR operation are comparable. The aforementioned findings also indicate that VGA successfully explores and fuses the information of graph and visual modalities.

\subsubsection{Similarity Measurement Module Replacement Experiment}

To further explore the relationship between features of propagation structures and other modalities. we replaced the cosine similarity in the similarity measurement module with alternative methods to compare their effectiveness.

\begin{table}[htbp]
  \centering
  \setlength{\tabcolsep}{13pt}
  %\small
  \caption{Similarity measurement module replacement experiment results.}
  \scalebox{1.0}{
  \begin{threeparttable}
    \begin{tabular}{c|cc|cc|cc}
    \toprule
    \multirow{2}[2]{*}{Method}  &\multicolumn{2}{c|}{WeiboCED}         & \multicolumn{2}{c|}{DRWeiboMM} & \multicolumn{2}{c}{Twitter} \\
     \cmidrule{2-7}  & \makecell[c]{Acc.} &  \makecell[c]{F1} & \makecell[c]{Acc.}  & \makecell[c]{F1} & \makecell[c]{Acc.} &  \makecell[c]{F1} \\
    \midrule
    VGA w/o Sim  & 93.48	&92.94	&86.24	&85.02	&83.78	&82.10 \\
    VGA-Euc  &93.88	&93.32	&86.72	&85.69	&83.86	&82.44 \\
    VGA-MSE  & 94.54	&94.09	&87.14	&86.02	&84.42	&83.24 \\
    VGA & 95.17	&94.78	&88.37	&87.39	&85.87	&84.61 \\
    \bottomrule
    \end{tabular}%
    \end{threeparttable}
    }
  \label{tab4}
\end{table}%

The results are shown in Table~\ref{tab4}, where VGA-Euc uses Euclidean distance for similarity measure. VGA-MSE directly compares the mean square error between two vectors ($Z_{g}^{i}$ and $Z_{v}^{i}$) and replaces the similarity loss with the MSE loss. The results demonstrate that all three methods can exploit the relationship between the propagation structure and the visual modality, thereby enhancing the model's performance. However, compared to the other two methods, Euclidean distance exhibits a comparatively smaller improvement, which could be attributed to its limitations in handling high-dimensional, nonlinear, and non-Gaussian distributed data. Additionally, the cosine similarity method proves more suitable for this study's task than the mean square error method.

\subsubsection{Multimodal Fusion Replacement Experiment}

\begin{table}[htbp]
  \centering
  \setlength{\tabcolsep}{13pt}
  %\small
  \caption{Multimodal fusion replacement experiment results.}
  \scalebox{1.0}{
  \begin{threeparttable}
    \begin{tabular}{c|cc|cc|cc}
    \toprule
    \multirow{2}[2]{*}{Method}  &\multicolumn{2}{c|}{WeiboCED}         & \multicolumn{2}{c|}{DRWeiboMM} & \multicolumn{2}{c}{Twitter} \\
     \cmidrule{2-7} & \makecell[c]{Acc.} &  \makecell[c]{F1} & \makecell[c]{Acc.}  & \makecell[c]{F1} & \makecell[c]{Acc.} &  \makecell[c]{F1} \\
    \midrule
    VGA-Concat  & 94.61	&94.16	&87.67	&86.37	&84.29	&82.53 \\
    VGA-Weight  &94.79	&94.35	&87.76	&86.62	&84.54	&82.76 \\
    VGA-SA  & 94.99	&94.67	&87.9	&86.87	&84.64	&82.79 \\
    VGA & 95.17	&94.78	&88.37	&87.39	&85.87	&84.61 \\
    \bottomrule
    \end{tabular}%
    \end{threeparttable}
    }
  \label{tab5}
\end{table}%

This subsection explores the effect of various modal fusion methods on performance, and the experimental results are shown in Table~\ref{tab5}. Where VGA-Concat directly concatenates visual features and graph features as fused features. VGA-Weight employs a weighted summation method for fusing each modal feature. VGA-SA concatenates the features of the two modalities and uses the self-attention mechanism for high-level feature extraction. According to Table~\ref{tab5}, VGA-Concat underperforms as it does not employ any modal fusion methods. By contrast, VGA-SA and VGA outperform VGA-Weight, demonstrating the efficacy of the attention mechanism. Notably, VGA attains the highest performance, demonstrating the superiority of the co-attention mechanism in modal fusion.

\begin{figure}
\centering
\includegraphics[width=0.65\textwidth]{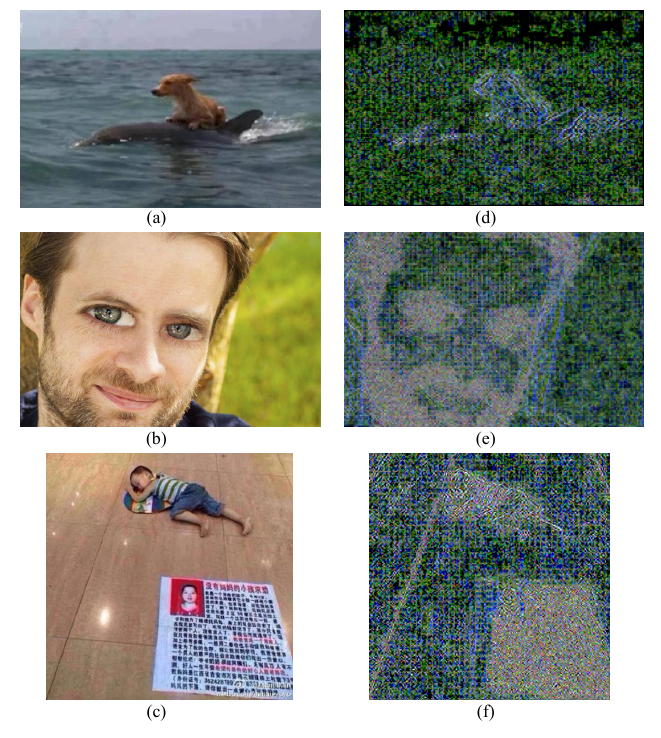}
\caption{Some rumors detected by VGA but missed by VGA w/o Noise. Where (a) (b) (c) are the original RGB images in the datasets, and (d) (e) (f) are the transformed noise images.}
\label{fig5}
\end{figure}

\subsection{Case Study}

To better demonstrate the importance of noise images for rumor detection, we compare the results of VGA and its variant VGA w/o Noise, demonstrating instances in which VGA can correctly predict but VGA w/o Noise cannot in Figure~\ref{fig5}.

As for the three claims in Figure~\ref{fig5}, VGA w/o Noise incorrectly predicts their veracity due to the fact that they all describe an event with a small impact and a high probability of occurring in reality. Consequently, there are fewer contradictory and questionable comments in their responses, and there are no obvious traces of manipulation in the RGB image, resulting in detection errors. However, after added image tampering features, VGA is able to make accurate predictions, demonstrating the effectiveness of image tampering feature extraction.

\section{Conclusions}

In this work, we propose a novel Vision and Graph Fused Attention Network (VGA) for rumor detection, which can better utilize propagation structures while further exploring visual features and similarity relationships between modalities. We analyze the characteristics of current rumor data and introduce image noise information to extract tampering features. Embed textual, propagation structure and image features of rumors are integrated in an unified framework. 
%And the correlations between the features of propagation structures and other modalities is fully utilized by the similarity measurement module.
%In visual information processing, VGA captures both semantic and tampering information of the images and employs an attention mechanism to produce higher-order feature representations. In Graph information processing, VGA constructs each claim as an isomorphic graph and improves its interaction with the root node in order to extract spread structure features more efficiently.
Empirical studies demonstrate that VGA is more accurate than the state-of-the-art methods on all three real-world datasets.

% \section*{Limitations}
% Although our model outperforms the state-of-the-art methods on small-scale datasets like Twitter. However, the improvement is modest when compared to the outcomes of other datasets. VGA still needs a certain size dataset to improve the accuracy of rumor classification in order to meet the desired practical application criteria.

\begin{acks}
The authors would like to thank all the anonymous reviewers
for their help and insightful comments. This work is supported
in part by the National Natural Science Foundation of China (61876016), the National Key R\&D Program of China (2018AAA0100302) .
\end{acks}

% \clearpage
\bibliographystyle{ACM-Reference-Format}
\bibliography{sample-base}

\end{document}